\documentstyle[prl,multicol,aps,amstex]{revtex}
\begin{document}

\input{epsf.tex}
\epsfverbosetrue

\title{Pulse Shepherding and Multi-Channel Soliton Transmission in
Bit-Parallel-Wavelength Optical Fiber Links}

\author{Yuri S. Kivshar and Elena A. Ostrovskaya}

\address{Optical Sciences Centre, Australian Photonics Cooperative Research Centre, Australian National University \\ Canberra
ACT 0200, Australia}

\maketitle
\begin{abstract}
We study bit-parallel-wavelength (BPW) pulse transmission in multi-channel
single-mode optical fiber links for high-performance computer networks. We
develop a theory of the pulse shepherding effect earlier discovered in
numerical simulations, and also describe families of the BPW solitons
and bifurcation cascades in a system of N coupled nonlinear
Schr\"odinger equations.
\end{abstract}
\begin{multicols}{2}
\narrowtext
\vspace{-5mm}
A growing demand for high-speed computer communications requires an
effective and inexpensive parallel computer interconnects that eliminate bottlenecks caused by parallel-to-serial conversion. Recently, bit-parallel-wavelength (BPW) single-fiber optical links were proposed as possible high-speed computer interconnects \cite{review}. Functionally, the BPW link is just a single-medium parallel fiber ribbon cable. In a conventional ribbon cable, $N$ parallel bits coming from a computer bus are transmitted by $N$ pulses travelling in $N$ separate fibers. In a BPW scheme, all $N$ bits are wavelength multiplexed and transmitted by time-aligned pulses, ideally -- solitons, through a {\em single optical fiber}. 

For any bit-parallel transmission the crucial problem is mantaining the alignment of pulses corresponding to parallel bits of the same word. Unlike the fiber ribbon cable, a single-fiber BPW link presents a unique possibility of {\em dynamical control} of the pulse alignment by employing the so-called {\em pulse shepherding} effect \cite{wid,yeh1}, when a strong "shepherd" pulse enables manipulation and control of co-propagating weaker pulses. Experimentally, the reduction of the misalignment due to the group-velocity mismatch of two pulses in the presence of the shepherding pulse has been observed in a dispersion-shifted Corning fiber \cite{exp}. 

In this Letter, we develop a rigorous theory of the shepherding effect, and show that it is caused by the nonlinear cross-phase modulation (XPM) of pulses transmitted through the same fiber. The co-propagating pulses can be treated as fundamental modes of different colour trapped and guided by an effective waveguide induced by the the strong shepherd pulse. The resulting {\em multi-component soliton pulse} propagates in the fiber preserving the time alignment of its constituents and thus enables multi-channel bit-parallel transmission. For the first time to our knowledge, we analyze multi-component solitons and describe a mechanism of the soliton multiplication via the bifurcation cascades in a model of $N$ nonintegrable coupled nonlinear Schr\"odinger (NLS) equations.

To describe the simultaneous transmission of $N$ pulses of different wavelengths in a single-mode optical fiber, we follow the standard derivation \cite{hasegawa} and obtain a system of $N$ coupled NLS equations in the coordinate system moving with the group velocity $v_{g0}$ of the central pulse $(0 \leq j \leq N-1)$: 
\begin{equation}\label{a_eq}
i \left(\frac{\partial}{\partial z} + \delta_j \frac{\partial}{\partial t}\right)A_j=
\frac{\alpha_{j}}{2} \frac{\partial^2 A_j}{\partial t^2} - \gamma_j \left(|A_j|^2 + S_{mj}\right) A_j,
\end{equation}
where $S_{mj}=2 \sum_{m \neq j}^{N-1} (\gamma_m/\gamma_j)|A_m|^2$. 
For a pulse $j$, $A_j(z,t)$ is the
slowly varying envelope measured in the units of $\sqrt{P_0}$, where $P_0$ is the incident power carried by the central pulse, $\alpha_j=-\beta_{2j}/|\beta_{20}|$ is the normalized group velocity dispersion, $\delta_j=(v_{g0}-v_{gj})/v_{g0}v_{gj}$ is the relative group velocity mismatch, and $\gamma_j=\omega_j/\omega_0$ characterises the nonlinearity strength $(\alpha_0=\gamma_0=1)$. The time, $t=(T-Z/v_{g0})/T_0$, and propagation distance, $z=Z/L_0$, are measured in units of characteristic pulse width, $T_0\approx 10\,\,ps$, and dispersion length $L_0=T^2_0/|\beta_{20}|\approx 50\,\, km$ \cite{yeh1,exp}. For the operating wavelengths spaced $4\div5$ $nm$ apart (to avoid the four-wave-mixing effect), within the band $1530\div 1560$ $nm$ (see Refs. \cite{yeh1,exp}), the coefficients $\alpha_j$ and $\gamma_j$ are different but close to $1$. For the realistic group-velocity difference of less than $5$ $ps/km$ \cite{exp}, the mismatch parameter $\delta_j \leq 1$. 

Below we show that the system (\ref{a_eq}) admits stationary solutions in the form of multi-component {\em BPW solitons} which represent a shepherd pulse time-aligned with a number of lower-amplitude pulses. We then discuss the effect of the group-velocity mismatch on the time alignment of the constituents of the multi-component pulse. 

To find the stationary solutions of Eqs. (\ref{a_eq}), we use the transformation: $A_j(t,z)=u_j(t)\exp(i \delta_j \alpha_j t + i \lambda_j z)$, with the amplitudes $u_j$ obeying the following equations:
\begin{eqnarray}
\label{NLS_Nn}
\frac{1}{2} \frac{d^2 u_0}{dt^2} + \left(u_0^2 +
2\sum_{n=1}^{N-1}\gamma_n u_n^2 \right) u_0
= \frac{1}{2} u_0,
\nonumber \\
\frac{\alpha_n}{2}\frac{d^2u_n}{dt^2} + \gamma_n \left( u_n^2 + 2\sum_{m
\neq n}^{N-1} \frac{\gamma_m}{\gamma_n}u_m^2 \right) u_n = \lambda_n u_n,
\end{eqnarray}
where the amplitudes, time, and $\lambda_n$ are measured in units of $\sqrt{2\lambda_0}$, $(2\lambda_0)^{-1/2}$, and $2\lambda_0$, respectively.

System (\ref{NLS_Nn}) has {\em exact analytical solutions} for $N$ coupled
components. Indeed, looking for solutions
in the form $u_0(t) = U_0 {\rm sech} \, t$, $u_n(t) = U_n {\rm
sech} \; t$, we obtain the constraints $\lambda_n = \alpha_n/2$, and a system
of $N$ coupled algebraic equations,
\[
U_0^2 + 2 \sum_{n=1}^{N-1}\gamma_n U_n^2 = 1,   \;\; \gamma_n U_n^2 + 2 \sum_{m \neq n}^{N-1}
\gamma_m U_m^2 = \alpha_n.
\]
As long as all modal parameters, $\alpha_n,\,\gamma_n$, are close to $1$, this solution describes a composite pulse with $N$ nearly equal constituents. 
In the degenerate case, $\alpha_n = \gamma_n =1$, the amplitudes are \cite{yeh5}: $U_0 = U_n = [1 + 2(N-1)]^{-1/2}$. 

Approximate analytical solutions of different types can be obtained in the {\em linear limit} \cite{man}, when the central frequency pulse, $u_0$, is much larger than other pulses, and the XPM interaction between the latter can be neglected.
Lineariziation of Eqs. (\ref{NLS_Nn}) for $|u_n| \ll |u_0|$ yields an exactly solvable NLS equation, for $u_0$, and $N-1$ decoupled linear equations, for $u_n$. Each of the latter possesses a
localized solution provided $\lambda_n = \Lambda_n \equiv (\alpha_n/8) [1 - \sqrt{1 +
16\alpha_n}]^2$. Near this point, the central soliton pulse (shepherd pulse) can be thought of as inducing an effective waveguide that supports a fundamental mode $u_n$, with the corresponding cutoff, $\Lambda_n$. Since $\alpha_n$
and $\gamma_n$ are close to $1$, the soliton-induced
waveguide always supports no more than two modes {\em of the same wavelength}, with largely separated eigenvalues. As a result, the effective waveguide induced by the shepherd pulse stays predominantly {\em single-moded} for all operating wavelengths.

Let us describe the mechanism of pulse shepherding in more details. First, we consider the simplest case $N=2$. If the two pulses, $(0)$ and $(1)$, do not interact, then the {\em uncoupled} Eqs. (\ref{NLS_Nn}) possess only single-component soliton solutions $u_0(t) = {\rm
sech} \,  t$ and $u_1=(2\lambda_1\gamma_1/\alpha_1)^{1/2}{\rm sech} \sqrt{2 \lambda_1/\alpha_1} t$, with the corresponding normalized powers $P_0 \equiv \int u^2_0 dt = 2$ and $P_1= 2(2 \lambda_1 \gamma_1)^{1/2} \alpha^{-3/2}_1$. These solutions can be characterised by the curves on the diagram $P(\lambda)$ [see curves $(0)$ and $(1)$ in Fig. 1]. When the two copropagating pulses interact, a new branch of the {\em two-mode} solitons $(0+1)$ appears (branch A-B in Fig. 1). It is characterised by the total power $P(\lambda_1)=P_0+P_1$, and it joins the two branches $P_0(\lambda_1)$ and $P_1(\lambda_1)$ at the {\em bifurcation points} $O_1$ and $O_2$, respectively. Near the point $O_1$, the solution consists of a large pulse of the central wavelength that guides a small component $u_1$ (see Fig. 1, point $A$). The point $O_1$ therefore coincides with the cutoff $\lambda_1=\Lambda_1$ for the fundamental mode $u_1$ guided by the shepherd pulse $u_0$. Shapes and amplitudes of the soliton components evolve with changing $\lambda_1$ (see the point $B$ in Fig. 1). The 
component $u_0$ disappears at the bifurcation point $O_2$. 

Next, we consider a
shepherd pulse guiding three pulses, $n=1,2,3$, with the modal parameters: $\alpha_{0,3} = \gamma_{0,3} =
1$, $\alpha_1=\gamma_1=0.65$, and $\alpha_2 =\gamma_2=0.81$. Solitary waves
of this {\em four-mode BPW system} can be found as localized solutions of Eqs.
(\ref{NLS_Nn}), numerically.  Figure 2 presents the lowest-order families of such
localized solutions on the line $\lambda_1=\lambda_2=\lambda_3\equiv \lambda$ in the parameter space $\{\lambda_1,\lambda_2,\lambda_3\}$. If the pulses $(1)$, $(2)$, and $(3)$, were interacting only with the central pulse but not with each other, then the bifurcation pattern for each of this pulses would be similar to that shown in Fig 1. Thin dotted, dashed, and dash-dotted curves in Fig. 2 correspond to the solitons of three independent pulses $(1)$, $(2)$, and $(3)$, shown with branches of corresponding two-mode solitons of the BPW system with pairwise interactions, $(0+1)$,  
$(0+2)$, and $(0+3)$, respectively (cf. Fig. 1). In fact, all four pulses interact with each other, and therefore each new constituent added to a multi-component pulse ``feels'' the effective potential formed not only by the shepherd pulse but also by all the weaker pulses that are already trapped. In addition, mutual trapping of the pulses $(1)$, $(2)$ and $(3)$ without the shepherd pulse is possible. As a result, new families of the two-mode $(1+2)$ and, branching off from it, the three-mode $(0+1+2)$ solutions appear (marked curves in Fig. 2).  The three-mode solutions bifurcate at the point $O_3$ and give birth to the four-mode $(0+1+2+3)$ solitons (branch $O_3-O_4$). An example of such four-wave composite solitons is shown in Fig. 2 (inset). This solution corresponds to the typical shepherding regime of the BPW transmission for $N=4$, when the central pulse $u_0$ traps and guides three smaller fundamental-mode pulses on different wavelengths. 

On the bifurcation diagram (Fig. 2), starting from the central pulse branch, the solution family undergoes a {\em cascade of bifurcations}: $O_1\to O_2 \to O_3 \to O_4$. On each segment of the corresponding solution branches, different multi-component pulses are found: $(0) \rightarrow (0+1) \rightarrow (0+1+2) \rightarrow (0+1+2+3) \rightarrow (1+2+3)$. The values of the modal parameters in Fig. 2 are chosen to provide a clear bifurcation picture, although they correspond to the wavelength spacing that is much larger than the one used in the experiments \cite{exp}, for which $\gamma_n/\gamma_{n+1}\approx 0.997$. If the modal parameters are tuned closer to each other, the first two links of the bifurcation cascade tend to disappear. Ultimately, for equal parameters, the bifurcation
points $O_2$ and $O_3$ merge at the point $O_1$, and the four-mode soliton family (thick line in Fig 2) branches off directly from the
central pulse branch $(0)$. Then, near the point $O_1$, the four-mode pulse can be described by the linear theory. The
qualitative picture of the {\em bifurcation cascade} for $N=4$ preserves for other values of $N$. 

The BPW solitons supported by shepherd pulses are {\em linearly stable} \cite{dima}. However, the effect of the relative walk-off due to the group velocity mismatch of the soliton constituents endangers the integrity of a composite soliton and thus the pulse alignment in the BPW links.  It is known that, in the case of two solitons of comparable amplitude,  nonlinearity can provide an effective trapping mechanism to keep the pulses together \cite{menyuk}. In
the shepherding regime, a multi-component pulse creates an effective attractive potential, and the $j$-th pulse is trapped if its group velocity is less than the {\em escape velocity} of this potential. The threshold value of the walk-off parameter can be estimated as \cite{hasegawa}: $\delta^2_j\leq (4 \alpha_j/P_j)\sum_{m\neq j} \gamma_m u^2_m(0)$, where $u^2_m(0)$ is the peak intensity of the component $m$.  For instance, for the component $j=1$ of the four-component BPW soliton presented in Fig 2 (point A), the estimated threshold $\delta_1\leq 1.7$ agrees with the numerically calculated value $\delta_1\leq 2.2$. 

In reality, all the components of a BPW soliton would have nonzero walk-off. The corresponding numerical simulations are
presented in Fig. 3 for $N=4$. Initially, we launch
four pulses as an exact four-mode BPW soliton $A$ (see Fig. 2) of Eqs. (\ref{NLS_Nn}). When this soliton evolves along the fiber length, $z$, in the presence of small
to moderate relative walk-off ($\delta_j \neq 0$ for $j\neq 0$), its components remain strongly localized and mutually trapped [Figs. 3(a,b)], whereas it loses more 
energy into radiation for much larger values of the relative walk-off [Figs. 3(c,d)]. The former situation is more likely to be realized experimentally as the relative group-velocity mismatch for pulses of different wavelength is different \cite{exp}. In this case, the conclusive estimate for the threshold values  of $\delta_j$ can only be given if the shepherd pulse is much stronger than the guided pulses, which are approximately treated as non-interacting fundamental modes of the effective waveguide induced by the shepherd pulse. 

In conclusion, we have developed a theory of the shepherding effect in BPW fiber links and established that the pulse shepherding can enable the time alignment of the copropagating pulses despite the relative group velocity mismatch.

The authors acknowledge fruitful discussions with A. Hasegawa,  C. Yeh and L. Bergman and a partial support of the Performance and Planning Fund.

\begin{figure}
\setlength{\epsfxsize}{6.0cm}
\centerline{\epsfbox{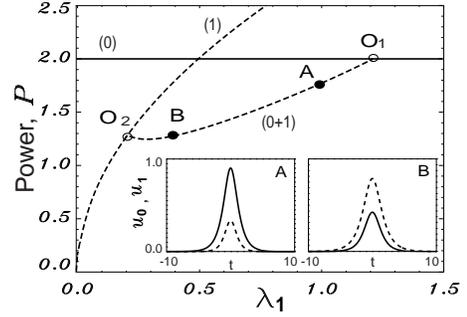}}\vspace{3mm}
\caption{Bifurcation diagram of the two-channel BPW model for $\alpha_{0,1}=\gamma_{0,1}=1.0$: horizontal line - branch of the central pulse (0); dashed - branches of the (1) and (0+1) solitons ($A-B$). Inset: examples of the (0+1) solitons corresponding to the points $A, \,B$, shown for the $u_0$ (solid) and $u_1$(dashed) components.}
\label{fig1}
\end{figure}
\begin{figure}
\setlength{\epsfxsize}{6.0cm}
\centerline{\epsfbox{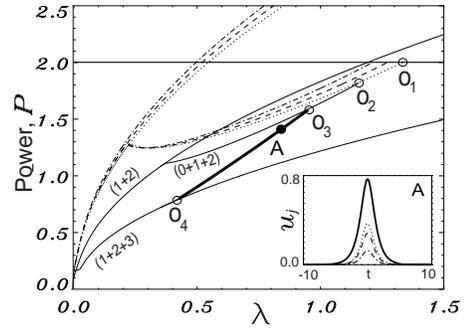}}\vspace{3mm}
\caption{Bifurcation diagram of the four-channel BPW model: horizontal line - branch of the central pulse (0); dotted, dashed and dash-dotted - branches of pulses (1), (2), and (3) with the (0+1), (0+2), and (0+3) solitons, respectively; solid thin - branches  of the (1+2), (0+1+2), and (1+2+3) solitons; thick - branch of the (0+1+2+3) solitons.  Inset: example of the (0+1+2+3) soliton corresponding to the point $A$; solid, dotted, dashed and dash-dotted lines - amplitudes of the $u_0$, $u_1$, $u_2$, and $u_3$ components, respectively.}
\label{fig2}
\end{figure}
\begin{figure}
\setlength{\epsfxsize}{6cm}
\centerline{\epsfbox{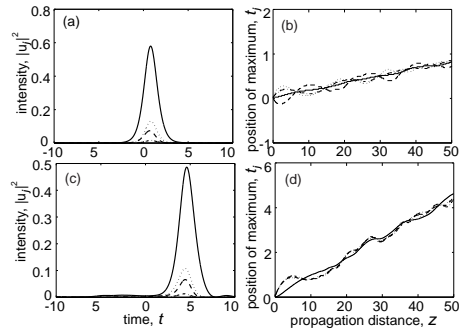}}\vspace{3mm}
\caption{Influence of the mode walk-off on the shepherding effect for the input parameters as for soliton $A$ of Fig. 2. Shown are: components of the output pulse at $z=50$ for (a) $\delta_1=0.45, \delta_2=-0.35, \delta_3=0.25$, and (c) $\delta_1=\delta_2=\delta_3=0.9$; (b,d) - evolution of the position of the maxima for all pulse constituents. Solid line - maximum of the shepherd pulse.}
\label{fig3}
\end{figure}
\end{multicols}
\end{document}